# Functionalized Carbon Nanotube Electrodes for Controlled DNA Sequencing

Rameshwar L. Kumawat[†], Biswarup Pathak*[,†,#]

[†]Discipline of Metallurgy Engineering and Materials Science, and [#]Discipline of Chemistry, School of Basic Sciences, Indian Institute of Technology (IIT) Indore, Indore, Madhya Pradesh, 453552, India

*E-mail: biswarup@iiti.ac.in

**Abstract**

In the last decade, solid-state nanopores/nanogaps have attracted significant interest in the rapid detection of DNA nucleotides. However, reducing the noise through the controlled translocating of the DNA nucleobases is a central issue for the developing nanogap/nanopore-based DNA sequencing to achieve single-nucleobase resolution. Furthermore, the high reactivity of the graphene pores/gaps exhibits clogging of the pore/gap, leading to the blockage of the pores/gaps, yielding sticking, and irreversible pore closure. To address the prospective of functionalization of carbon nanostructure and for accomplishing this objective, herein, we have studied the performance of functionalized closed-end cap carbon nanotube (CNT) nanogap-embedded electrodes which can improve the coupling through nonbonding electrons and may provide possibility of N/O-H···π interaction with nucleotides, as single-stranded DNA is transmigrated across. We have investigated the effect of functionalizing the closed-end cap CNT (6,6) electrodes with purine (adenine, guanine) and pyrimidine (thymine, cytosine) molecules. Weak hydrogen bonds formed between the probe molecule and target DNA nucleobase enhance the electronic coupling and temporarily stabilize the translocating nucleobase against the orientational fluctuations, which may reduce noise in the current signal during experimental measurements. The findings of our density functional theory and non-equilibrium Green's function-based study indicates that this modeled setup could allow DNA nucleotide sequencing with a better and reliable yield, giving current traces that differ by at least 1 order of current magnitude for all four target nucleotides. Thus, we feel that functionalized CNT nanogap-embedded electrodes may be utilized for controlled DNA sequencing.

## 1. Introduction

Recent advances in DNA sequencing have paved the role in personalized medicine, which is the next frontier in our health care, as it could be used to detect predisposition concerning several genetic illnesses, and finally delivers accurate treatments.[1-4] To fully accomplish this, enhanced control and cost of the procedure are further needed to be improved.[4-5] The progress towards faster, reliable, and cheaper sequencing has been very demanding since the "$1000 Human Genome Project" launched.[3]

Nanopore/nanogap based human genome sequencing is one of the most developing technologies since it is promising to provide faster, reliable, and cheaper sequencing. Furthermore, it has the potential to bring genomic science into personalized medicine.[1-10] In the last decades, significant development and successes have been made. In 2007, Jin He and co-workers suggested the functionalization of gold (Au) nano-electrodes with cytosine probe/reader for DNA (deoxyribonucleic acid) sequencing.[11] After that, in 2008, Haiying He and co-workers have tested four DNA nucleobase molecules as probes and found cytosine probe to yield robust results in terms of nucleobase distinguishability.[12] Again, in 2009, Jin He and co-workers have demonstrated that the tunneling with double-functionalized contacts could read the nucleobase structure of the unmodified DNA oligomers with a resolution analogous to that of ion-current read-outs in the nanopores/nanogaps.[13] In 2012, Pathak and co-workers investigated the effect of chemically double-functionalization on Au nano-electrodes for enhancing the nanopore-based DNA sequencing.[14] They have demonstrated that the chemically functionalized molecular probes are capable of temporarily forming hydrogen (H) bonds with the incoming DNA base part and the phosphate group. Thus, reducing the noise and further slowdown the translocating speed of the nucleobases between the Au nanoelectrodes. Furthermore, Su and co-workers have reported that guanine-functionalized Au nanoelectrodes revealed promising results.[15] However, transverse conductance across the DNA molecules located between the two Au nano-electrodes has been aggressively examined and debated. Therefore, special attention has been devoted to exploring low-dimensional (2D and 1D) materials.[14-16]

Low-dimensional materials for example graphene,[1,2,4-8,10,17-27] hexagonal-boron nitride (hBN),[30,31] graphene-hBN,[32] molybdenum disulfide ($MoS_2$),[33,34] silicene,[35] black phosphorene,[36,37] and various other low dimensional materials have been studied for DNA sequencing owing their excellent electronic and transport properties. For the progress of pore/gap based DNA sequencing techniques, all these materials have provided new paradigms since their atomic thickness can be comparable to the DNA nucleobase spacing in single-stranded DNA (ssDNA).[20,24,] Mostly, graphene-based nanopore/nanogap,[2,24] nanoribbon,[38] and carbon nanotubes (CNTs)[16] have been explored as potential nano-electrodes materials for DNA sequencing. So far, several experimental and theoretical works have done on DNA sequencing using graphene electrodes.[20,24,22,29] Traversi and co-workers have experimentally demonstrated that the solid-state nanoscale-sized pore integrated with the graphene nanoribbon could be utilized for single-nucleobase sensing.[27] However, DNA passing through graphene-based pores/gaps confirmed, a significant drawback of the technique was that the DNA nucleobase adhered to the nanopore/nanogap surface. Hence, the reactivity of the graphene edge leads to geometry deformation.[24] Such reactivity also leads to the clogging of the pore, leading to the blockage of the nanopores, yielding sticking and irreversible pore closure.[2,24,39] Several theoretical works have also reported that precise functionalization of such pores/gaps or nonperforated regions with ligands such as aryl-, or alkyl- groups may, though, minimize the clogging of DNA nucleotides and deliver better surface energy, as well as diffusion performance.[40,41] Prasongkit and one of the present authors have tried improving the sensitivity by chemically functionalizing both the graphene electrodes to improve the electronic coupling with the translocating target DNA base.[42] Despite all these works, experimental and theoretical researchers are continually searching for a low-dimensional nanopore/nanogap based electrode material that can provide molecular-level resolution with enhanced control.[2,16,24,32]

Low-dimensional CNTs have an atomically thin structure (chemical and physical) and unique electrical transport properties,[43-49] which can allow the individual identification of DNA nucleotides based on in-plane transverse transmission and current signals and may accomplish the single-nucleobase resolution with enhanced control. Thus, to realize the potential of solid-state DNA sequencing on low-dimensional materials and be able to read nucleotides at a single-molecular level, also, to enhance the readout time, a high-fidelity control of ssDNA translocation still needs to be accomplished. Motivated by these reports, we predict that CNT can be a better electrode as they may not be as reactive as the graphene edges, and the presence of π-clouds in CNT may improve the coupling with the translocating DNA nucleotides. Specifically, such curvature (benzene-like six-membered rings at the closed-end cap provides the possibility of π-π coupling with nucleotides) in the CNT may improve the N/O-H··· π interaction with the DNA nucleotides. Additionally, suitable functionalizing groups can be attached to the CNT caps, which could lead to a better electronic coupling with the incoming ssDNA nucleobases through the formation of temporarily H-bonds.[11-15,40,41] Such a functionalization molecular probe could be attached to only one electrode or both electrodes. Functionalization of the CNTs caps could be a better way to achieve single-nucleobase resolution, reducing the noise in transverse electric current signals and further slowdown the translocating speed of the target DNA nucleobases between the functionalized CNT electrodes. Also, functionalized CNT-based nanogap could simplify the DNA sequencing process and allows different types of electrical measurements (such as conductance and transverse electronic current). That being the case, CNT is very promising and opens possibilities for controlled translocation of DNA nucleotides. Hence, the present study purposes of exploring the applicability of functionalized closed-end cap CNT (6,6)-based nanogap-embedded electrodes for DNA sequencing.

In this work, we have studied the structural, electronic, and transport properties of the four setups, namely deoxyadenosine monophosphate (dAMP), deoxyguanosine monophosphate (dGMP), deoxythymidine monophosphate (dTMP), and deoxycytidine monophosphate (dCMP), when inserted inside a functionalized CNT (6,6)-based nanogap. This has been done to accomplish an unambiguous distinction of all four nucleotides in DNA sequencing. Electronic structure and binding energies are analyzed using the density functional theory (DFT). Further, current-Voltage ($I - V$) characteristic, zero-bias transmission function, and bias-dependent transmission function for all four nucleotides analyzed using the non-equilibrium Green's functional combined with DFT.

## 2. Model and Computational Methods

**Figure 1** shows the proposed nanogap device consists of two semi-infinite functionalized closed-end cap CNT (6,6)-electrodes with a DNA nucleotide. Here, we have taken purine or pyrimidine types DNA nucleobase molecule as the reader molecule, acting as the molecular probe for the functionalized CNT (6,6) electrodes. We have examined all four molecular probes for sequencing the DNA nucleobases [**Figure 1; Figure S1 (Supporting Information)**].

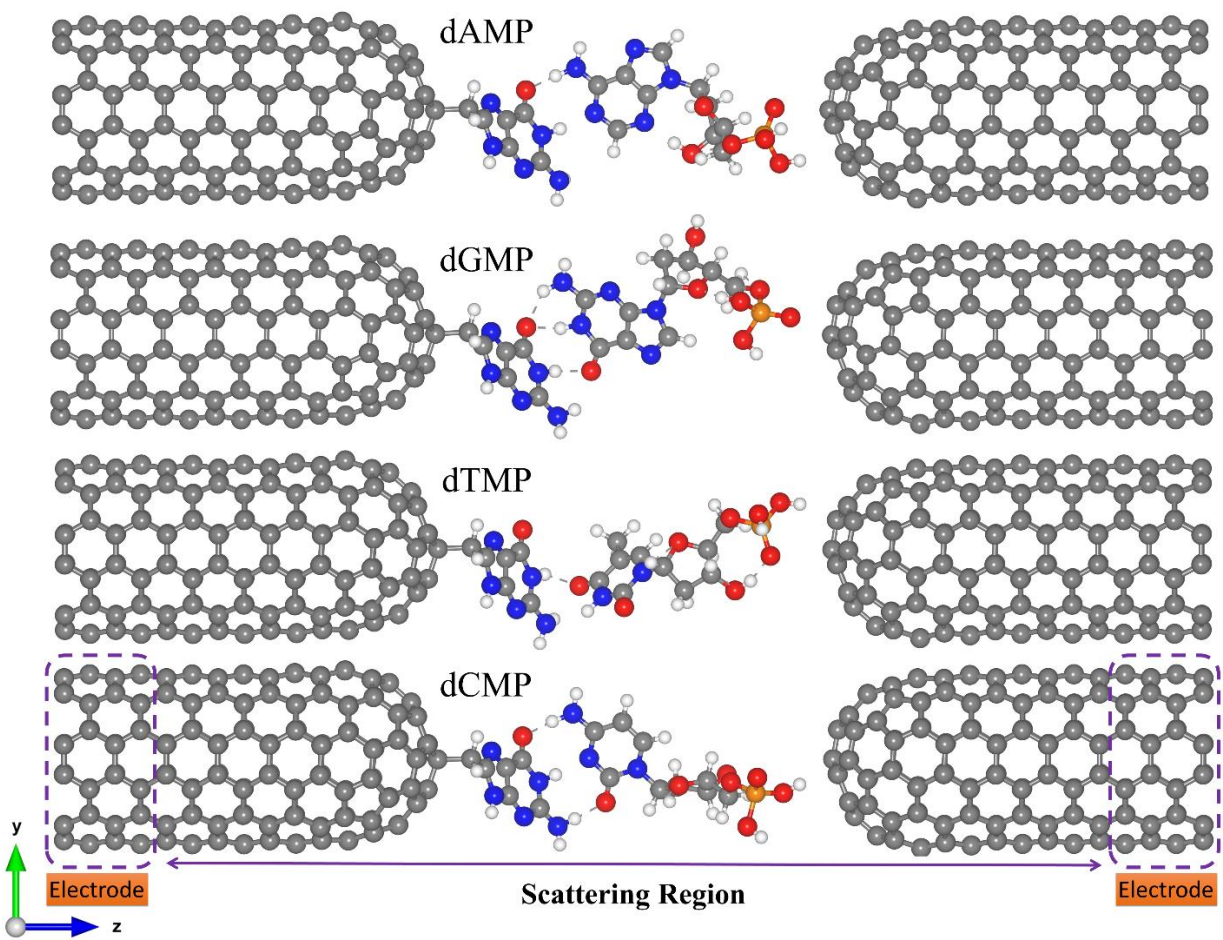

**Figure 1**. Atomic structure of the proposed functionalized closed-end cap CNT (6,6)-based nanogap setup for the detection of four different nucleotides (dAMP, dGMP, dTMP, and dCMP). The CNT (6,6)-electrodes (left and right) are semi-infinite and periodic along the transport direction ($z$−axis). Here, the CNT cap is functionalized by a guanine nucleobase. Atom color code: P (orange), O (oxygen), N (blue), C (grey), H (white).

The pairing geometry between the reader molecule and the target nucleotide molecule (a target is a segment of a ssDNA including nucleobase, sugar, and phosphate group) is obtained by performing full geometry relaxation of an isolated pair of two nucleobases using the DFT methodology with B3LYP/6-31G* level of theory as implemented in the Gaussian 09 code.[50] This is done to get the intermolecular binding configurations and then proceeded to place each base pair in the nanogap. The electrode-electrode spacing of 19.01 Å (shortest gap length between two CNT electrodes) is maintained throughout the calculations. Both the semi-infinite electrodes are periodic in the $z-$direction. We have considered a significant vacuum distance (along $x$ and $y-$direction) of 28 Å, which is enough to avoid any unphysical interaction between the repetitive images of the nanogap system.

The whole nanogap setup (Figure 1) is then fully optimized by employing the DFT methodology as implemented in the SIESTA code.[51,52] We have used GGA-PBE (generalized gradient approximation with Perdew-Burke-Ernzerhof) approximation for exchange and correlation functional.[53] Norm-conserved Troullier-Martins pseudopotentials are used to describe the interactions between the core and valence electrons.[54] We have used the mesh cut-off value of 200 Ry for real space integration and DZP (double-zeta polarized) basis sets, including polarization orbitals for all atoms.[53,32-37] We have considered Γ-point for the sampling of the Brillouin zone, due to the large cell size. Structures are relaxed by the conjugate-gradient (CG) algorithm using the tolerance in density matrix difference is 0.0001, and the atomic forces are lesser than 0.01 eV/Å.

The binding energy ($E_b$) is calculated using the following equation:

$$E_b = [E_{CNT+DNA} - (E_{CNT} + E_{DNA})] \quad (1)$$

where $E_{CNT+DNA}$ represents the total optimized energy of the pristine/functionalized CNT+DNA nucleotide setup. Here $E_{CNT}$, $E_{DNA}$ are the energy of the pristine/functionalized CNT setup and DNA nucleotide molecule, respectively within the geometry of the pristine/functionalized CNT+DNA nucleotide setup.

The electronic transport properties have been carried out using the Landauer-Buttiker approach. We have used the NEGF technique combined with DFT (NEGF+DFT), as implemented in the TranSIESTA code.[51,55] The basis set and the real-space integration used in the electronic transport calculation is the same as used for the geometry relaxation. The transverse electric current is calculated according to the following equation:

$$I(V_b) = \frac{2e}{h} \int_{\mu_R}^{\mu_L} T(E, V_b) [f(E - \mu_L) - f(E - \mu_R)] dE \qquad (2)$$

where $T(E, V_b)$ represents the transmission function of the electrons entering at energy ($E$) from $L$ to $R$ electrode due to applied bias voltage ($V_b$), $f(E - \mu_{L,R})$ represents the Fermi-Dirac distribution of electrons in the $L/R$ electrodes, and $\mu_{L,R}$ is the chemical potential here $\mu_{L/R} = E_F \pm V_b/2$ is moved respectively up/down, according to the Fermi energy $E_F$.[13,14,32-37]

## 3. Results and Discussion

As the ssDNA is translocated through the functionalized CNT-based pore/gap by a driving electric field, the probing molecule (A/G/T/C) will interact with each nucleotide while translocating over the pore/gap setup through forming weak H-bonds with the reader nucleobase and nucleobase part of the target DNA nucleotide molecule. The DNA nucleobases appearing in a natural DNA have a characteristic capability to bind to their corresponding complementary nucleobase molecules selectively.[11-14] Hence, we have considered all four nucleobase molecules (i.e., A, G, T, and C) as the reader/probe molecules in our study. Firstly, we have compared the interaction strength of the DNA nucleobases towards the electrode. The interaction strength between the CNT (pristine/functionalized) and the target DNA nucleotide molecules investigated by analyzing the binding energy ($E_b$) as defined in equation 1 (**Model and Computational Methods**). The calculated binding energy values are given in **Table S1**. From **Table S1**, we have found that functionalized CNT-based nanogap has high $E_b$ values compared to pristine CNT-based nanogap, which is important for electrical measurements. Further, the $E_b$ values of the reader-target nucleobase pairs formed temporarily in the nanogap is investigated for all four readers-target nucleotide systems. The interaction between O and H bond is stronger than that of N and H because of the higher electronegativity of an oxygen atom. The $E_b$ of dGMP is higher than that of dAMP because dGMP involved in four hydrogen bonding with two oxygen atoms (**Figure S2,b**). In the case of dCMP, the higher $E_b$ is due to three hydrogen bonding (2 O-H and 1 N-H; **Figure S2,d**). We compared dAMP and dTMP; dAMP is forming two hydrogen bonds (O-H and N-H; **Figure S2,a**) while dTMP is forming only one hydrogen bond (O-H; **Figure S2,c**); therefore, dTMP shows lower $E_b$. The $E_b$ value of the reader-target nucleobase pairs formed temporarily in the nanogap can also be correlated to the number of H-bonds formed in between the reader-target base pairs. For example, in the case of guanine-guanine base pair the $E_b$ is -0.58 eV per H-atom. Nevertheless, the formation of H-bond is subjected to the orientations of the target molecule. The calculated $E_b$ values show that the formation of temporarily H-bonding can stabilize the DNA molecules for a short time as it passes through the nanogap. Thus, H-bonds formed between the reader molecule and target DNA nucleobase enhance the electronic coupling and stabilize the translocating nucleobase against the orientational fluctuations and thus may significantly reduce noise in the transverse electronic current signal.[11-14,42] From calculated $E_b$ values, we have perceived that guanine-reader gives better electronic coupling compared to other readers (A/T/C) and pristine CNT. This indicates that the guanine-reader could be a better reader molecule for controlled and rapid DNA sequencing. Further, from the point of view of a DNA sequencing device, we have studied the I-V characteristics and transmission function for all four nucleotides detection using the NEGF+DFT approach.

Next, we investigated the current-voltage ($I - V$) characteristic for the nanogap setup using four different probes (**Figure 1** and **Figure S1**). The figure of merit for identifying all four DNA nucleotides is that their accompanying currents (*I*) should be differed by at least 1 order of current magnitude. We have tabulated our recognition chart in **Table 1**. For example, at a bias voltage of 0.10 V, using either C or T as a probe, we can distinguish the set dAMP and dGMP from dCMP and dTMP (dAMP from dGMP and dCMP from dTMP however cannot be distinguished from each other)) (**Table 1, Figure S3**). The A probe offers different currents for dAMP, dGMP or dCMP and dTMP (dAMP and dGMP can be distinguished while dCMP and dTMP though remain indistinguishable from each other). In contrast, when base G is used as a probe, we can distinguish dAMP, dTMP, and dCMP from dGMP (while dAMP, dTMP, and dCMP cannot be distinguished from each other) (**Table 1; Figure 2**).

**Table 1**. Summary of the genetic information deducible for the different probes (i.e., A, G, T, and C) from current measurement at different applied bias voltages (at 0.10 V, 0.50 V, and 0.70 V). Target nucleobases that cannot be distinguished are separated by (,).

| probe | V= 0.10 V | V= 0.50 V | V= 0.70 V |
|---|---|---|---|
| A | dAMP \| dGMP \| dCMP, dTMP | dAMP, dGMP \| dCMP \| dTMP | dAMP, dGMP \| dCMP \| dTMP |
| G | dAMP, dCMP, dTMP \| dGMP | dAMP \| dGMP \| dTMP \| dCMP | dAMP \| dGMP \| dTMP \| dCMP |
| T | dAMP, dGMP \| dCMP, dTMP | dAMP, dGMP, dTMP, dCMP | dAMP, dCMP, dTMP \| dGMP |
| C | dAMP, dGMP \| dCMP, dTMP | dAMP, dCMP \| dGMP, dTMP | dAMP, dGMP, dTMP, dCMP |

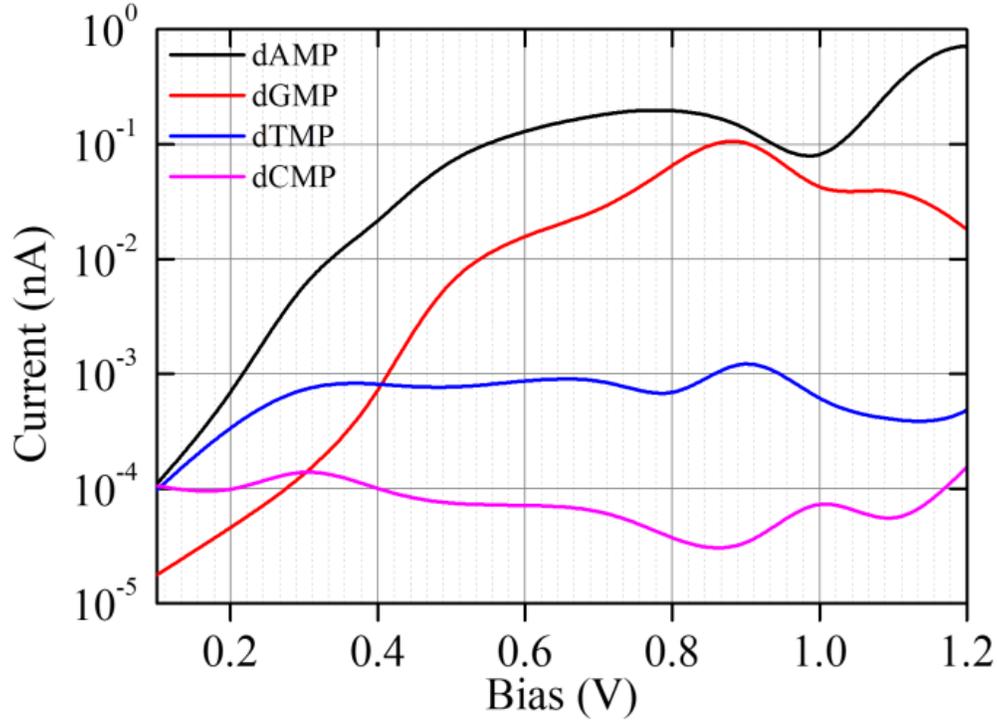

**Figure 2.** The $I-V$ characteristic curves (plotted on a semi-logarithmic scale) for the nanogap setup functionalized with a G-probe for the four different target nucleotides (dAMP, dGMP, dTMP, and dCMP).

It is observed that at an increased applied bias voltage, the detection properties of the nanogap setup change. For example, at 0.50 V the A probe delivers different current signals, and we can distinguish the set dAMP and dGMP from dCMP and dTMP (dCMP and dTMP can be distinguished while dAMP and dGMP, however, remains indistinguishable from each other). Using C probe, we can distinguish dGMP and dTMP from dAMP and dCMP (dGMP from dTMP and dAMP from dCMP, however, cannot be distinguished from each other). Moreover, using the T probe, we cannot distinguish all four DNA nucleotides inside the gap. Furthermore, at 0.70 V, using the C probe, we can distinguish dAMP, dGMP, and dTMP from dCMP (while dAMP, dGMP, and dTMP cannot be distinguished individually). Using the T probe, we can distinguish dAMP, dCMP, and dTMP from dGMP (while dAMP, dCMP, and dTMP cannot be distinguished individually). However, using the A probe, we can distinguish dAMP and dGMP from dCMP and dTMP (dCMP and dTMP can be distinguished while dAMP and dGMP, however, remains indistinguishable from each other).

We have found that the functionalization of the CNT-electrodes with G probe gives a unique identification of the four DNA nucleotides (**Figure 2**). This can be accomplished by performing three sequencing runs at different applied bias voltages (i.e., at 0.10 V, 0.50 V, and 0.70 V). We have presented a flowchart (**Figure 3)**, which gives the illustration of individual identification of all four target nucleotides in DNA. At 0.10 V would results in a sequence of current traces that categories into two effortlessly distinguishable trends known as high (H) and low (L) current values. The higher current values if A (i.e., dAMP) or C (i.e., dCMP) is the target nucleotide and low current values if G (i.e., dGMP) or T (i.e., dTMP) is the target nucleotide. Herein, we have noted that at 0.10 V dGMP can easily identify; however, the other three nucleotides cannot be distinguished. Therefore, we require a high bias voltage to resolve the remaining ambiguity of all four nucleotides in DNA. At 0.50 V, it will be possible to distinguish between the purine (dAMP, dGMP) and pyrimidine (dCMP, dTMP) type's nucleotides (**Figure 2**). The difference between the two types of nucleotides is around 1 orders of current magnitude, which should make the distinction strong. Moreover, our proposed systems show the ability to identify dAMP and dGMP nucleotide, which is very important for nanogap DNA sequencing. Hence, at 0.50 V, all four nucleotides become distinguishable with their respective current traces. Besides, the current difference of all four nucleotides (dAMP, dGMP, dTMP, and dCMP) is more than 1 order of current magnitude. Similarly, at 0.70 V also provides the current signals for the nucleotides dAMP, dGMP, dTMP,

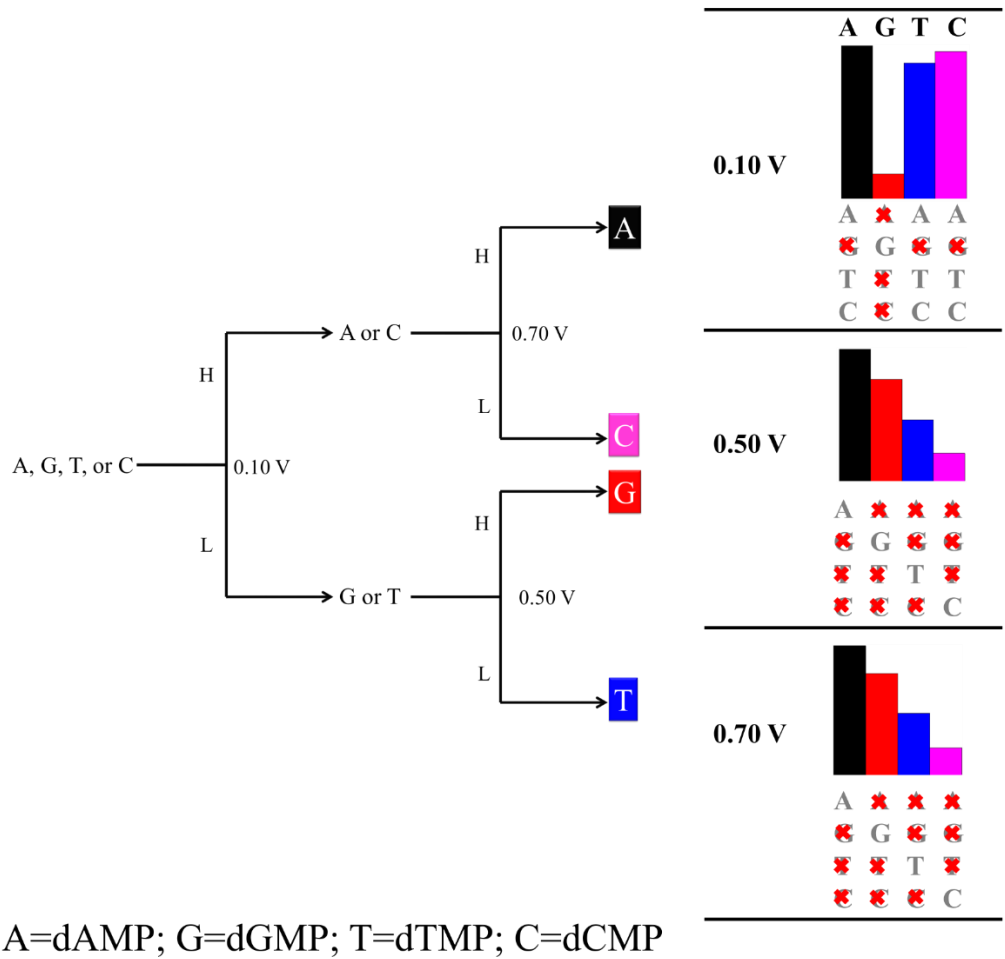

A=dAMP; G=dGMP; T=dTMP; C=dCMP

**Figure 3**. Flowchart illustrating the decision-making process of a nanogap setup involving nucleobase guanine as a probe leading to the identification of each target nucleotide in the DNA sequence. Herein, High (H) and Low (L) refer to higher and lower current values at a given bias voltage (V). The height of the bars below the letters A, G, T, and C on the right side of the figure corresponds to the respective current traces. The crossed-out letters below the bars refer to possible target nucleotides that have been ruled out.

and dCMP to differ by at least 1 order of current magnitude. This is leading to an easy distinction between the four nucleotides, where the high current values correspond to dAMP nucleotide, while the low current values correspond dCMP nucleotide. More precisely, dAMP and dCMP gives the highest and lowest current signals, whereas dGMP and dTMP can be identified in between the former two nucleotides.

After demonstrating the most important ability of our proposed nanogap system for DNA sequencing, we study the zero-bias transmission function and bias dependent transmission for all four nucleotides.[12,14,32,36-38,42] Herein, we focus on the most promising nanogap setup involving the guanine base as a probe. **Figure 4(a/b)** shows the zero-bias transmission function together with the zero-bias density of states (DOS) when located inside the functionalized closed-end cap CNT (6,6) electrodes. It is noted that the DOS peak matches to the positions of the transmission function peaks. The associated molecular orbitals (MOs) with transmission peaks near the Fermi-level are presented in **Figure 4**. We have seen in Figure 4 that for the four DNA nucleotides, the Fermi-level is allied closely to the highest occupied molecular orbitals (HOMOs), in contrast, lowest unoccupied molecular orbitals (LUMOs) are allied far from the Fermi-level. In the near Fermi region, we have found that for the given nanogap setup the transmission function has a similar shape for the four nucleotides which can be due to the transmission peaks near the Fermi region and within the MO gap which is far from specific MOs are typically associated with the contact states localized on the guanine-probe and the CNT (6,6) electrodes (**Figure 4c**). The different nucleotide types influence the transmission function at zero-bias. The size of purines is larger than pyrimidines, and the number of hydrogen atoms available for interaction is also different in these two groups. Thus, the size of the target DNA nucleobase, thereby the gap between nucleobase and the CNT (6,6) electrode, decides the transmission magnitude through the overlap between the states localized on the nucleobase molecule and the states on the CNT (6,6) electrode. Consequently, from the transmission functions at zero-bias, one can conclude that the dissimilarity of physical and chemical structures between the purine and pyrimidine nucleotides affects the electronic coupling strength of the nucleotides with the CNT (6,6) electrodes. As a result, they are leading to the prospect of distinguishing the two different groups of DNA nucleotides under bias. The zero-bias transmission function and corresponding DOS using base adenine, cytosine, and thymine as a probe are presented in **Figure S4**.

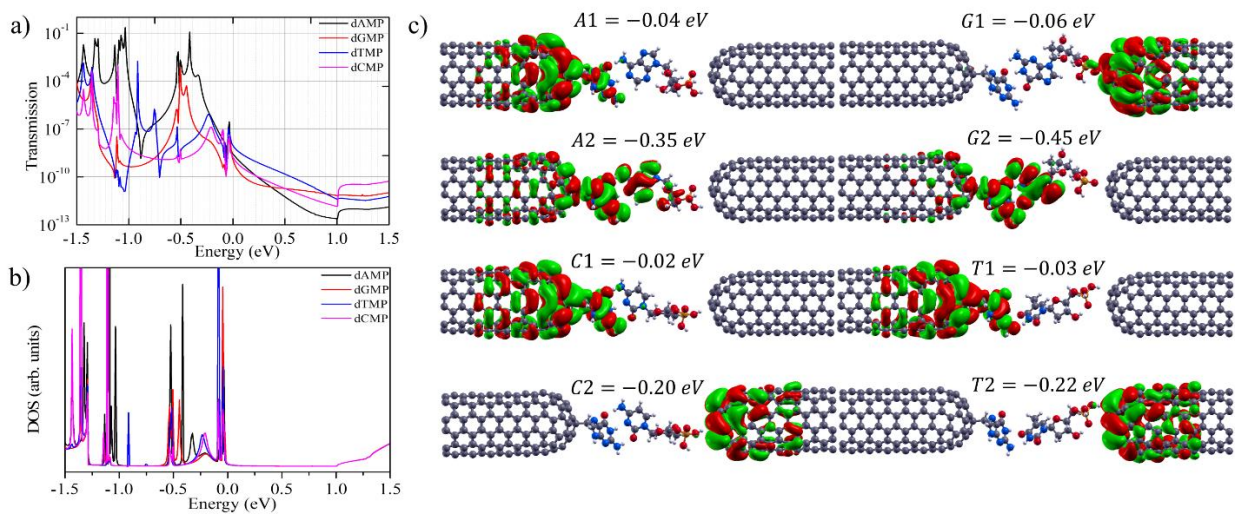

**Figure 4**. (a) Nanogap setup with guanine-probe. Zero-bias transmission function plotted on a semi-logarithmic scale for all four target nucleotides (dAMP, dGMP, dTMP, and dCMP). (b) Zero-bias DOS plotted for all four nucleotides. The Fermi-level has been aligned to zero energy. (c) Molecular orbitals responsible for those specific transmission peaks (with respective energies) are presented for all four target nucleotides.

Next, to achieve a deep understanding of our above discussed $I - V$ characteristic features, we have analyzed the bias-dependent transmission function for the four target DNA nucleotides while inserted inside the guanine-probe functionalized CNT (6,6) electrodes (**Figure 5**). It is important to note that under applied bias, the system is in the non-equilibrium state; hence, it has less relation with binding energy. Therefore, once the system is in the non-equilibrium state, the most important parameter is the movement of the HOMO/LUMO peak associated with the electrode and molecule. That means molecular states present in the system will play a significant role in I-V signals.[12,14,19,32,36-38,42] This is the reason; we have explored the bias-dependent transmission function for all four nucleotides. Herein, when the bias is increased from 0 to 0.50 V, the transmission function peaks linked with the contact states enter the bias region and contribute to the increase of the current magnitude across the nanogap setup. The energy levels in the guanine-probe increase following the chemical potential of the left electrode as the bias is increased. Also, the position of the transmission peaks moves form high energy to low energy closely following the potential of the left electrode and right electrode. The HOMO (A1, T1, and C1) state of dAMP, dTMP and dCMP localized on the guanine-probe and the CNT (6,6) electrodes are closest to the Fermi-level which in result yields the highest current magnitude in the low bias region (0.1 to 0.2 V). Furthermore, the purine nucleotides (dAMP and dGMP), when compared to the pyrimidine nucleotide (dTMP and dCMP), the HOMO (A2 and G2) states of purine nucleotides localized on the nucleobase part as well as on the guanine-probe enter the bias region and contribute to the increase of the current magnitude across the nanogap setup. This could be a reason that purine nucleotides (dAMP and dGMP) affords higher current magnitude and pyrimidines (dTMP and dCMP) lower current magnitude with increased bias (0.4 to 1.2 V). This is very much in agreement with our $I-V$ curves too. Thus, we have concluded that molecular states on the nucleotides, and the localized states on the guanine-probe, and CNT (6,6) electrodes, play a significant role in the transmission. Depending upon the variation in the transmission function for the four DNA nucleotides, the trends in current follow specific variations in the lower and higher bias window, thus facilitating principle capability for unambiguous identification of all four DNA nucleotides.

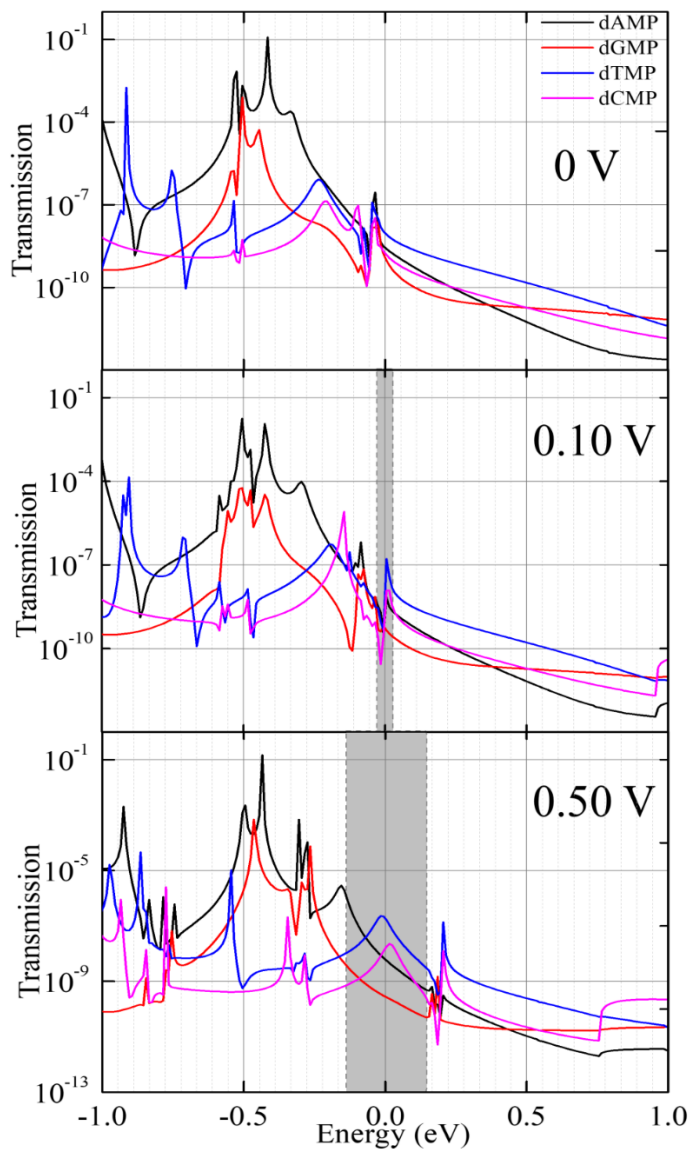

**Figure 5**. Nanogap setup with Guanine-probe. The bias-dependent transmission functions plotted on a semi-logarithmic scale for all four target nucleotides (dAMP, dGMP, dTMP, and dCMP), with variation of energy $E$ at different bias voltages (0 V, 0.10 V, 0.50 V).

## 4. Conclusion

In conclusion, we have shown that functionalized closed-end cap CNT (6,6)-based nanogap with the specific molecular probe could lead to an enhancement in the sensitivity of the target nucleobase detection in a nanogap-translocating DNA sequence. Through the formation of temporary H-bonds, the electronic coupling could be enhanced, and the incoming nucleotide would be temporarily stabilized between the two electrodes, which potentially allowing for less orientational fluctuations and thus reducing noise in the current signals. Further, functionalized closed-end CNT (6,6) nanogap is providing more time for each nucleobase and allows different types of electrical measurements. From the transmission function and the $I - V$ characteristics of the target nucleotides, we have shown that guanine-probe could lead to a significant improvement in the sensitivity of nucleotides detections in a nanogap-translocating DNA sequence. We find a considerable difference in the current signals for all four nucleotides is more than 1 order of magnitude at two different voltages (0.5 and 0.7 V). This leads to an easy distinction between purine and pyrimidine type nucleobases. More specifically, dAMP and dCMP affords higher and lower current signals, though dGMP and dTMP can be identified in between the former two nucleotides with around 1 order of current magnitude. Bias-dependent transmission function reveals the molecular states contributing to the $I - V$ signals, which play a significant role in the DNA detection process. This is because of the electronic coupling of the target nucleobase with these states, as well as the localized state on the guanine-probe and electrodes provides molecules characteristic as recorded through the $I - V$ curve. Therefore, we believe that functionalized closed-end cap CNT (6,6)-based nanogap electrodes may be utilized for controlled DNA sequencing.

## 5. Associated Contents

* Supporting Information

## 6. Conflicts of interest

There are no conflicts of interest to declare.

## 7. Acknowledgments

We thank IIT Indore for the lab and computing facilities. This work is supported by DST-SERB, (Project Number: EMR/2015/002057) New Delhi and CSIR [Grant number: 01(2886)/17/EMR (II)] and (Project Number: CRG/2018/001131) and SPARC/2018-2019/P116/SL. R. L. K thanks MHRD for research fellowships. We would like to thank Dr. Vivekanand Shukla for fruitful discussion throughout this work.

*Table of Content (TOC):*

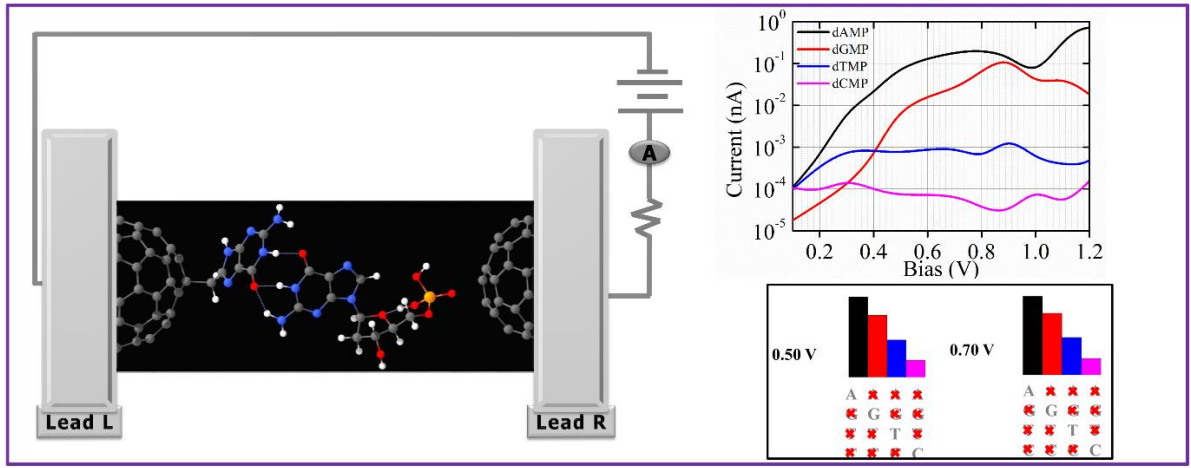